\documentclass[aps,preprintnumbers,nofootinbib,noshowpacs,eqsecnum,prd,superscriptaddress,letterpaper]{revtex4} 

\usepackage[utf8]{inputenc}
\usepackage{graphicx}
\usepackage{amsmath,amssymb}
\usepackage{url}
\usepackage{multirow}
\usepackage{hyperref,color}
\usepackage[normalem]{ulem}
\graphicspath{{figs/}}

\newcommand {\fbw}     {f_{BW}}
\newcommand {\fpone}   {f_{\Phi,1}}

\newcommand {\obw}     {{\cal O}_{BW}}
\newcommand {\opone}   {{\cal O}_{\Phi,1}}
\newcommand {\owww}    {{\cal O}_{WWW}}
\newcommand {\ow}      {{\cal O}_{W}}
\newcommand {\ob}      {{\cal O}_{B}}
\newcommand {\oww}     {{\cal O}_{WW}}
\newcommand {\obb}     {{\cal O}_{BB}}
\newcommand {\ogg}     {{\cal O}_{GG}}
\newcommand {\optwo}   {{\cal O}_{\Phi,2}}

\newcommand {\oqthree}  {{\cal O}^{(3)}_{\Phi,Q}}
\newcommand {\oqone}  {{\cal O}^{(1)}_{\Phi,Q}}
\newcommand {\our}  {{\cal O}^{(1)}_{\Phi,u}}
\newcommand {\odr}  {{\cal O}^{(1)}_{\Phi,d}}
\newcommand {\oud}  {{\cal O}^{(1)}_{\Phi,ud}}
\newcommand {\oer}  {{\cal O}^{(1)}_{\Phi,e}}
\newcommand {\ollll}  {{\cal O}_{LLLL}}
\newcommand {\ot}  {{\cal O}_{u\Phi,33}}
\newcommand {\otg}  {{\cal O}_{tG}}
\newcommand {\obo}  {{\cal O}_{d\Phi,33}}
\newcommand {\ota}  {{\cal O}_{e\Phi,33}}
\newcommand {\omu}  {{\cal O}_{e\Phi,22}}            
\newcommand {\fqthree}  {f^{(3)}_{\Phi,Q}}
\newcommand {\fqone}  {f^{(1)}_{\Phi,Q}}
\newcommand {\fur}  {f^{(1)}_{\Phi,u}}
\newcommand {\fdr}  {f^{(1)}_{\Phi,d}}

\newcommand {\fer}  {f^{(1)}_{\Phi,e}}
\newcommand {\fllll}  {f_{LLLL}}

\begin{document}

\title{Impact of CDF-II measurement of  $M_W$
  on the electroweak legacy of the LHC Run II}
\author{ Eduardo da Silva Almeida}
\email{eduardo.silva.almeida@usp.br}
\affiliation{Instituto de F\'isica, Universidade de S\~ao Paulo,
  R. do Mat\~ao 1371, 05508-090 S\~ao Paulo, Brazil}
\author{Alexandre Alves}
\email{aalves@unifesp.br}
\affiliation{Departamento de F\'isica, Universidade Federal de S\~ao Paulo, UNIFESP,
Diadema, S\~ao Paulo, Brazil}
\author{Oscar J. P. \'Eboli}
\email{eboli@if.usp.br}
\affiliation{Instituto de F\'isica, Universidade de S\~ao Paulo,
  R. do Mat\~ao 1371, 05508-090 S\~ao Paulo, Brazil}
\author{M.~C.~Gonzalez--Garcia}
\email{maria.gonzalez-garcia@stonybrook.edu}
\affiliation{Departament  de  Fisica  Quantica  i  Astrofisica
 and  Institut  de  Ciencies  del  Cosmos,  Universitat
 de Barcelona, Diagonal 647, E-08028 Barcelona, Spain}
\affiliation{Instituci\'o Catalana de Recerca i Estudis Avancats (ICREA)
Pg. Lluis  Companys  23,  08010 Barcelona, Spain.}
\affiliation{C.N. Yang Institute for Theoretical Physics, Stony Brook University, Stony Brook NY11794-3849,  USA}
%

\begin{abstract}
We analyze the impact of the recently released CDF-II measurement of
$W$ mass on the SMEFT analyses of the electroweak precision data  as
well as Higgs and electroweak diboson productions. We work in the 
Hagiwara, Ishihara, Szalapski, and Zeppenfeld  basis in  which
eight generation-independent operators enter in the
electroweak precision data at tree level and, unlike in the Warsaw basis,
the analysis of that set of data constrains all the eight Wilson
coefficients, without the need of combination with Higgs or
electroweak diboson data results. We show that in the global
analysis the determination of the coefficients of all operators which do not
enter the electroweak precision data are barely affected by the
new $M_W$ determination.

\end{abstract}

\maketitle


Recently the CDF collaboration released a new measurement of the $W$
mass~\cite{CDF:2022hxs}
\begin{equation}
  M_W = 80.4335 \pm 0.0094 \hbox{ GeV} \;,
\label{eq:mwcdf}
\end{equation}
that we incorporate in our analyses of the electroweak legacy of the
LHC Run II~\cite{Almeida:2021asy}; further details of our analyses can
be found in this reference. Here, we do not take the new $M_W$ result
at face value due to the tension between it and the previous
measurements. Instead, we consider the {\sl standard 
average}~\cite{deBlas:2022hdk}
\begin{equation}
  M_W = 80.413 \pm 0.008 \hbox{ GeV}\;.
\label{eq:mwsav}
\end{equation}
Moreover, taking into account that the top quark mass has a large
impact in the corrections to the precision electroweak observables we
also update the average top quark mass through the inclusion of the
new CMS determination~\cite{CMS:TOP-20-008}. Again, performing the 
standard average we consider the SM predictions as obtained with
~\cite{deBlas:2022hdk}
\[
  m_t = 171.79 \pm 0.38\hbox{ GeV}\;.
\]
Except for these two changes, our analyses proceed as in
Ref.~\cite{Almeida:2021asy}, {\em i.e.} we used the same datasets and
the Hagiwara, Ishihara, Szalapski, and Zeppenfeld (HISZ)
basis~\cite{Hagiwara:1993ck, Hagiwara:1996kf}.  In fact the HISZ basis
is most suitable for the direct quantification of new results which,
as for the case of the new $M_W$ (and $m_t$) determination, affect
mainly the electroweak precision data (EWPD).  This is so because this
basis was chosen, precisely, to avoid blind directions in the EWPD
analysis~\cite{Corbett:2012ja}. Eight operators enter the EWPD at tree
level and the EWPD analysis constrains each of the eight Wilson
coefficients, without the need of combination with Higgs or
electroweak diboson data (EWDBD) results.  This is not the case for
other basis, for example the Warsaw basis, for which the EWPD analysis
can only limit Wilson coefficient combinations making it more indirect
to interpret the results and requiring a global analysis of EWPD in
combination with Higgs and EWPDB to quantify the impact on individual
coefficients.  Of course in the HISZ basis when performing the global
analysis, the fact that the allowed ranges of the coefficients of the
operators entering the EWPD have been modified can indirectly result
into modifications of the coefficients of operators entering only the
Higgs or EWDBD. However, as we will see, within the present precision,
this is a very small effect.\smallskip

\begin{figure}[h!]
  \centering
  \includegraphics[width=0.45\textwidth]{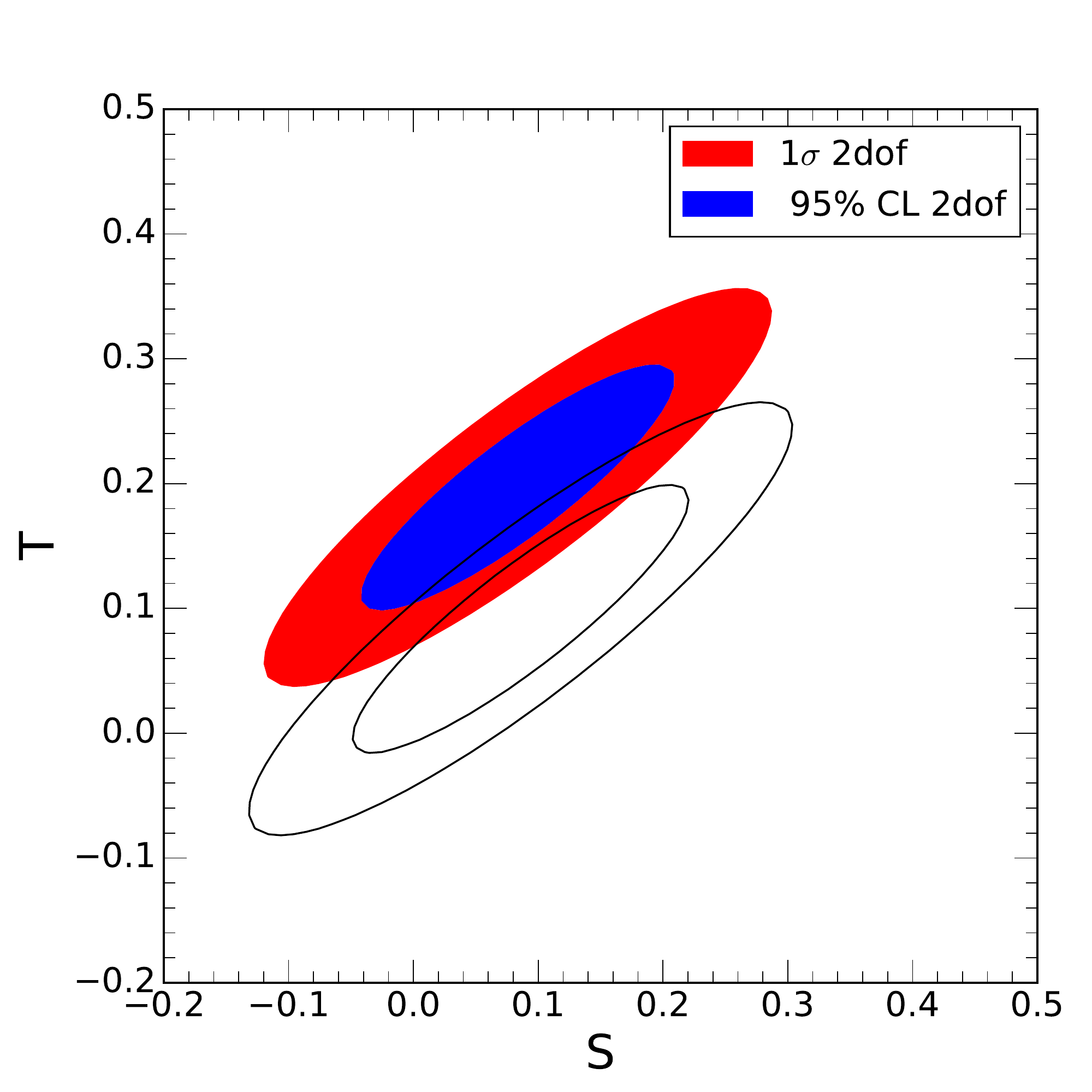}
  \caption{ $1\sigma$ and $2\sigma$ (2dof) allowed regions on the
    $ S\times T$ plane assuming that these are the only non-vanishing
    Wilson coefficients. The colored (transparent) region stands for
    the new (old) analysis. }
  \label{fig:2dfbwfpone}
\end{figure}

We start our analysis by considering a universal scenario where only
the $\fpone$ and $\fbw$ Wilson coefficients are non vanishing.
Figure~\ref{fig:2dfbwfpone} compares the $1\sigma$ and $2\sigma$
allowed regions in the plane $S \times T$
($S = -4\pi v^2\fbw/\Lambda^2$ and $ T=-2\pi v^2/e^2\fpone/\Lambda^2$)
obtained when using the previous and new values for $M_W$ and $m_t$.
As we can see, the allowed region using the new $M_W$ value is
displaced towards positive values with respect to the previous
result. In fact, $T$ is is $\simeq 3\sigma$ away from the SM
prediction ($T=0.197 \pm 0.065$) when we take into account the new
CDF-II result. We find a good agreement with the results in
Ref.~\cite{deBlas:2022hdk} performed under the same assumptions with
differences arising from the slightly different set of observables
considered. If instead of the average in Eq.~\eqref{eq:mwsav} one uses
the $W$ mass reported by CDF-II, Eq.~\eqref{eq:mwcdf}, the effect
would be obviously larger. This possible sign of new physics has
generated a cascade of analyses and models to explain
it~\cite{deBlas:2022hdk,Zhu:2022tpr,Fan:2022dck,Strumia:2022qkt,Athron:2022qpo,Yang:2022gvz,Tang:2022pxh,Du:2022pbp,Campagnari:2022vzx,Cacciapaglia:2022xih,Blennow:2022yfm,Sakurai:2022hwh,Fan:2022yly,Zhu:2022scj,Arias-Aragon:2022ats,Paul:2022dds,Babu:2022pdn,DiLuzio:2022xns,Bagnaschi:2022whn,Heckman:2022the,Cheng:2022jyi,Bahl:2022xzi,Song:2022xts,Asadi:2022xiy,Heo:2022dey,Crivellin:2022fdf,Endo:2022kiw,Du:2022brr,Cheung:2022zsb,DiLuzio:2022ziu,Balkin:2022glu,Biekotter:2022abc,Krasnikov:2022xsi,Zheng:2022irz,Ahn:2022xeq,Kawamura:2022uft,Peli:2022ybi,Ghoshal:2022vzo,Perez:2022uil}
    \smallskip

\begin{figure}[h!]
  \centering
  \includegraphics[width=0.75\textwidth]{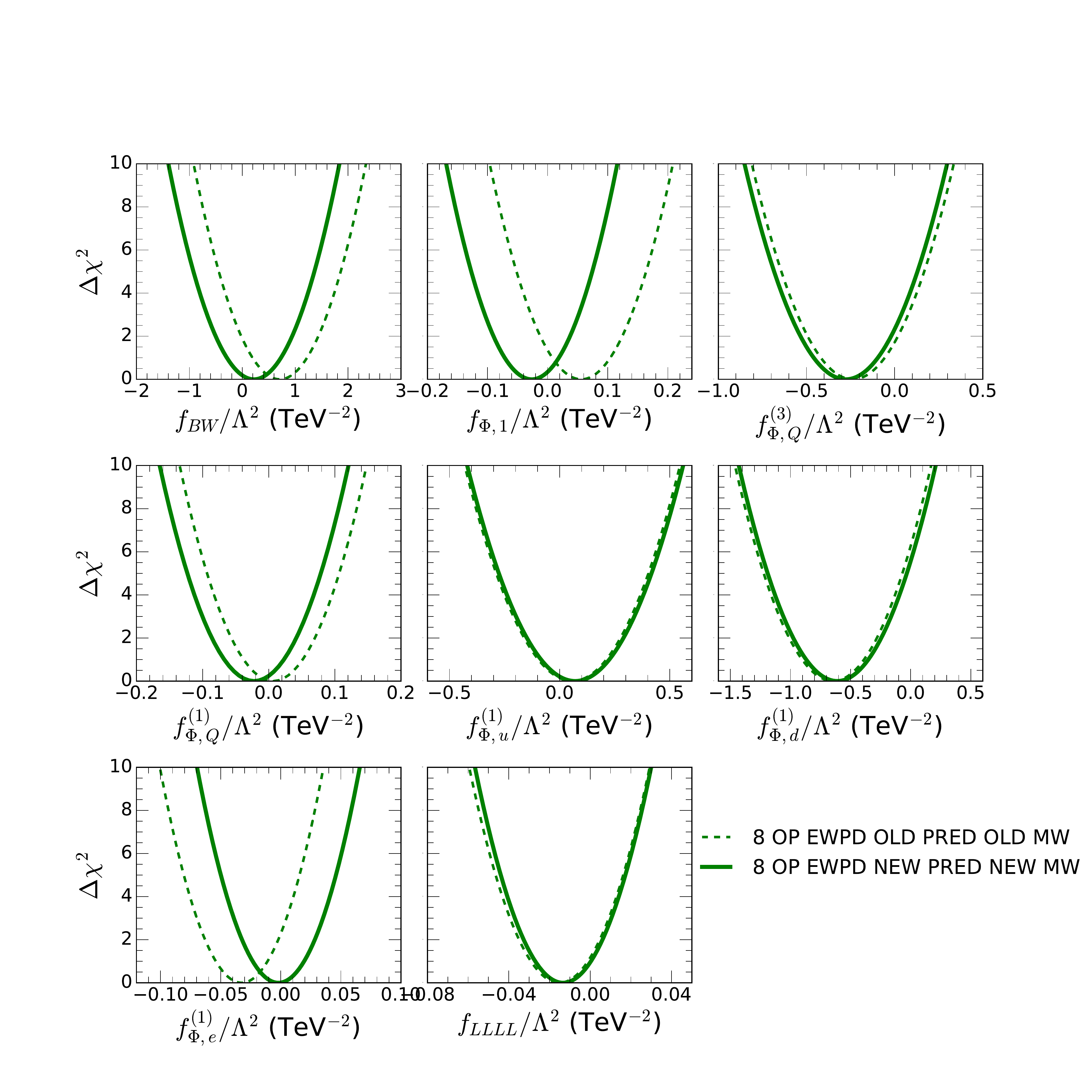}
  \caption{ $\Delta \chi^2$ as a function of the Wilson coefficients
    $\fbw/\Lambda^2$, $\fpone/\Lambda^2$, $\fqone/\Lambda^2$,
    $\fqthree/\Lambda^2$, $\fur/\Lambda^2$, $\fdr/\Lambda^2$,
    $\fer/\Lambda^2$, and $\fllll/\Lambda^2$, as indicated in the
    panels after marginalizing over the remaining fit parameters. The
    green solid (dashed) line stands for the fit of the EWPD
    with the new (previous) values of $m_t$ and $M_W$.}
  \label{fig:1dewpd}
\end{figure}

Next we perform the analysis in the full parameter space of 
SMEFT operators which affect the EWPD  at order $1/\Lambda^2$ in the HISZ
basis. We present in figure~\ref{fig:1dewpd} our results for this eight
parameter fit using the new $M_W$ and $m_{t}$ values (green line) compared to
those obtained with the previous values (dashed line). We find that
besides the effects on the oblique coefficients $\fbw$  and $\fpone$,  
the only coefficient of fermionic operators noticeably 
modified by the inclusion of the new data is $\oer$ (and also in less amount
$\fqone$) that contributes to the lepton
right-handed coupling. We notice, however, that the shift on the
fermionic operators originates
primarily from the new value for $m_t$ that modifies the SM predictions.
We notice also that the very small effect on the four-lepton operator
coefficient $f_{LLLL}$ is a consequence of the cancellation of the  variation
associated with the new $M_W$ and the new $m_t$. In summary, we find
that after including the eight relevant operators, the new value of $M_W$
still impacts mostly $\obw$ and $\opone$ but now in this 8-parameter analysis
their Wilson coefficients are still compatible with zero at 1$\sigma$.
Of course this occurs at the price of sizable correlations between the
allowed ranges of the eight coefficients. Altogether with find the
following best $\pm 1\sigma$ ranges and correlation matrix $\rho$
\begin{equation}
\begin{array}{c|r|rrrrrrrr}
{\rm Operator} &   \frac{f}{\Lambda^2} ({\rm Tev})^2 &
&&&\rho&&&  \\\hline
\obw     & 0.220\pm 0.511  & 1     &      &      &     &  & & & \\ 
\opone   &-0.026\pm 0.045  & 0.970 & 1    &      &     &  & & &\\
\oqone   &-0.022\pm 0.045  & 0.902 & 0.807& 1   &     &  &  & &\\
\oqthree &-0.277\pm 0.182  & 0.121 & 0.085& 0.152& 1  &  &  & &\\
\our     &0.070\pm 0.155  & 0.238& 0.185& 0.338 & -0.283& 1 & & &\\
\odr     &-0.611\pm0 .259 & 0.160& 0.166& 0.112& 0.605& -0.691& 1 &&\\
\oer     &-0.002\pm0.021 & -0.126 & -0.177& 0.033& 0.382& 0.580& -0.067& 1 & \\
\ollll &0.013\pm0.014   &  -0.952& -0.914& -0.851& -0.155& -0.293& -0.145&
0.052& 1\\
\end{array}
\end{equation}

\smallskip

\begin{figure}[h!]
  \centering
  \includegraphics[width=0.75\textwidth]{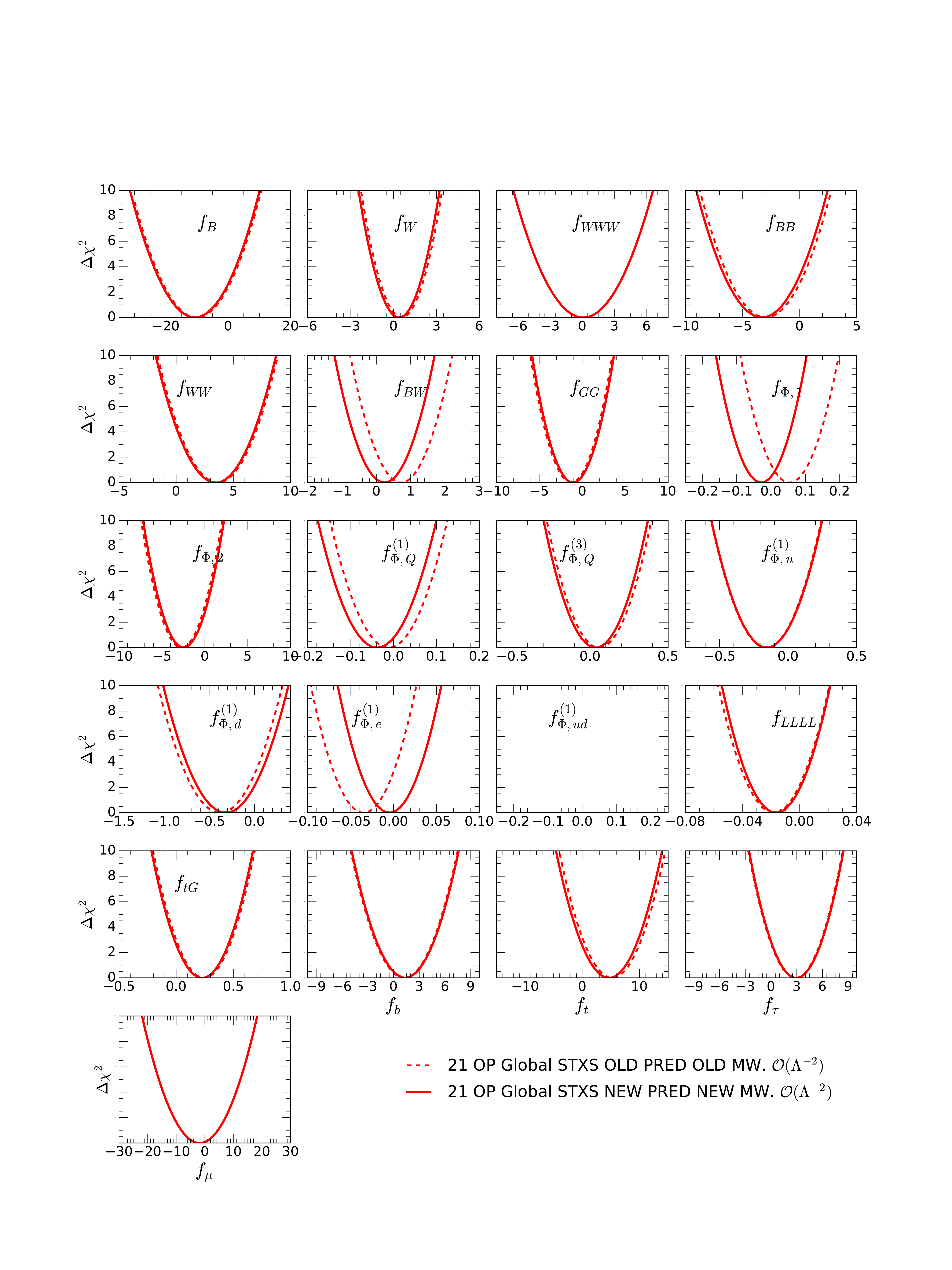}
  \caption{Marginalized one-dimensional $\Delta\chi^2$ distributions
    for the 21 parameters appearing in our global fit to EWPD, EWDBD,
    and Higgs data including the STXS Higgs data sets. The red solid
    (dashed) line stands for the results obtained with the new
    (previous) values for $m_t$ and $M_W$.  The theoretical
    predictions for the observables were evaluated at
    ${\cal O}(\Lambda^{-2})$ in the Wilson coefficients.}
  \label{fig:1dglobal}
\end{figure}

Finally we assess the impact of the new top and $W$ masses
on the global fit including EWPD and STXS-Higgs and diboson
datasets.
In figure~\ref{fig:1dglobal}, we depict the
marginalized one-dimensional $\Delta\chi^2$ distributions in the full
21-dimensional parameter space (of which only 20 coefficients
can be constrained at ${\cal O}(\Lambda^{-2})$).
The figure shows the results obtained including  the
new (solid line) compared to the previous (dashed line) values of $m_t$
and $M_W$. First of all, let us notice that there is no Wilson
coefficient that is incompatible with the SM predictions. This can be
seen clearly in Table~\ref{tab:newranges} that contains the 95\% CL
allowed regions for the operators entering our fit.
We also learn from figure~\ref{fig:1dglobal} that the Wilson
coefficients $\fbw$, $\fpone$ and $\fer$ are still the only ones which
are noticeably shifted while all others are barely modified.
\smallskip

\begin{table}
\begin{tabular}{|c||c|}\hline
  Operator & {95\% CL $f/\Lambda^2$ (TeV$^{-2}$)}\\
\hline
&  Global STXS  ${\cal O}(\Lambda^{-2})$, \\\hline
  $\ob$ 
& (-24,2.5) 
  \\
  $\ow$
  & (-1.4,2.2)
  \\
  $\owww$
  & (-4.1,4.2)
  \\
  $\obb$
  & (-6.9,0.34)
  \\ 
  $\oww$ 
  & (0.32,6.8)
  \\
  $\ogg$
  & (-4.1,1.9)
  \\
  $\otg$
  & (-0.053,0.51)
  \\
  $\optwo$
  &(-5.5,0.51)
  \\
  $\ot$
 & (-1.10,11)  
  \\
  $\obo$ 
  & (-2.6,5.3)  
  \\
  $\ota$ 
 & (-0.54,6.4)  
  \\
  $\omu$ 
  & (-15,11)
  \\
  $\obw$ 
  & (-0.68,1.2) 
  \\
    {$\opone$}
  & {(-0.13,0.05)}
    \\   
      {$\oqthree$}
    & (-0.17,0.25)  
    \\ 
      {$\oqone$}
    & {(-0.1,0.055)}
    \\
    $\our$
    & (-0.41,0.099)
    \\
    $\odr$
  & (-0.76,0.1
    \\
    $\oud$ 
    &  ---
    \\
    $\oer$
    &(-0.043,0.033)
    \\ 
    $\ollll$
    & (-0.041,0.0075)
    \\\hline
\end{tabular}
\caption{Marginalized 95\% CL allowed ranges for the Wilson coefficients
  global analyses including the STXS simplified template cross sections
  performed at at order ($1/\Lambda^2$).}
\label{tab:newranges}
\end{table}

\acknowledgments

This work is supported in part by Conselho Nacional de Desenvolvimento
Cent\'{\i}fico e Tecnol\'ogico (CNPq) grants 307265/2017-0 and
305762/2019-2, and by Funda\c{c}\~ao de Amparo \`a Pesquisa do Estado
de S\~ao Paulo (FAPESP) grants 2018/16921-1 and 2019/04837-9.  M.C.G-G
is supported by spanish grant PID2019-105614GB-C21 financed by
MCIN/AEI/10.13039/501100011033, by USA-NSF grant PHY-1915093, and by
AGAUR (Generalitat de Catalunya) grant 2017-SGR-929. The authors
acknowledge the support of European ITN grant
H2020-MSCA-ITN-2019//860881-HIDDeN.

\bibliography{references}
\end{document}